\documentclass[twoside,twocolumn,fleqn,superscriptaddress,showkeys]{revtex4-2}
\usepackage{xcolor}    
\usepackage{graphicx}  
\usepackage{amssymb}   
\usepackage{amsmath}   
\usepackage{bm}        
\usepackage{times}     
\usepackage{fancyhdr}  
\usepackage{hyperref}
\usepackage{url}
\begin{document} 

\title{The SABRE South Experiment at the Stawell Underground Physics Laboratory}

\author{I. Bolognino} 
\email{irene.bolognino@adelaide.edu.au}
\affiliation{Department of Physics, The University of Adelaide, Adelaide, SA 5005, Australia}
\affiliation{ARC Centre of Excellence for Dark Matter Particle Physics, Australia} 
\author{ for the SABRE South collaboration: E. Barberio}
\author{T. Baroncelli}
\affiliation{ARC Centre of Excellence for Dark Matter Particle Physics, Australia} 
\affiliation{School of Physics, The University of Melbourne, Melbourne, VIC 3010, Australia} 
\author{J. L. Bignell}
\affiliation{ARC Centre of Excellence for Dark Matter Particle Physics, Australia} 
\affiliation{Department of Nuclear Physics and Accelerator Applications, The Australian National University, Canberra, ACT 2601, Australia}
\author{I. Bolognino} 
\affiliation{Department of Physics, The University of Adelaide, Adelaide, SA 5005, Australia}
\affiliation{ARC Centre of Excellence for Dark Matter Particle Physics, Australia} 
\author{G. Brooks}
\affiliation{ARC Centre of Excellence for Dark Matter Particle Physics, Australia} 
\affiliation{School of Engineering, Swinburne University of Technology, Hawthorn, VIC 3122, Australia}
\author{F. Dastgiri}
\affiliation{ARC Centre of Excellence for Dark Matter Particle Physics, Australia} 
\affiliation{Department of Nuclear Physics and Accelerator Applications, The Australian National University, Canberra, ACT 2601, Australia}
\author{A. R. Duffy}
\affiliation{ARC Centre of Excellence for Dark Matter Particle Physics, Australia} 
\affiliation{Centre for Astrophysics and Supercomputing, Swinburne University of Technology, Hawthorn, VIC 3122, Australia}
\author{M. Froehlich}
\affiliation{ARC Centre of Excellence for Dark Matter Particle Physics, Australia} 
\affiliation{Department of Nuclear Physics and Accelerator Applications, The Australian National University, Canberra, ACT 2601, Australia}
\author{G. Fu}
\author{M. S. M. Gerathy}
\affiliation{ARC Centre of Excellence for Dark Matter Particle Physics, Australia} 
\affiliation{School of Physics, The University of Melbourne, Melbourne, VIC 3010, Australia} 
\author{G. C. Hill}
\affiliation{Department of Physics, The University of Adelaide, Adelaide, SA 5005, Australia}
\affiliation{ARC Centre of Excellence for Dark Matter Particle Physics, Australia}
\author{S. Krishnan}
\affiliation{ARC Centre of Excellence for Dark Matter Particle Physics, Australia} 
\affiliation{School of Engineering, Swinburne University of Technology, Hawthorn, VIC 3122, Australia}
\author{G. J. Lane}
\affiliation{ARC Centre of Excellence for Dark Matter Particle Physics, Australia} 
\affiliation{Department of Nuclear Physics and Accelerator Applications, The Australian National University, Canberra, ACT 2601, Australia}
\author{G. Lawrence}
\affiliation{ARC Centre of Excellence for Dark Matter Particle Physics, Australia} 
\affiliation{Centre for Astrophysics and Supercomputing, Swinburne University of Technology, Hawthorn, VIC 3122, Australia}
\author{K. T. Leaver}
\affiliation{Department of Physics, The University of Adelaide, Adelaide, SA 5005, Australia}
\affiliation{ARC Centre of Excellence for Dark Matter Particle Physics, Australia} 
\author{I. Mahmood}
\affiliation{ARC Centre of Excellence for Dark Matter Particle Physics, Australia} 
\affiliation{School of Physics, The University of Melbourne, Melbourne, VIC 3010, Australia} 
\author{P. McGee}
\affiliation{Department of Physics, The University of Adelaide, Adelaide, SA 5005, Australia}
\affiliation{ARC Centre of Excellence for Dark Matter Particle Physics, Australia} 
\author{L. J. McKie}
\author{P.C. McNamara}
\affiliation{ARC Centre of Excellence for Dark Matter Particle Physics, Australia} 
\affiliation{Department of Nuclear Physics and Accelerator Applications, The Australian National University, Canberra, ACT 2601, Australia}
\author{M. Mews}
\author{W. J. D. Melbourne}
\affiliation{ARC Centre of Excellence for Dark Matter Particle Physics, Australia} 
\affiliation{School of Physics, The University of Melbourne, Melbourne, VIC 3010, Australia} 
\author{G. Milana}
\affiliation{ARC Centre of Excellence for Dark Matter Particle Physics, Australia} 
\affiliation{School of Engineering, Swinburne University of Technology, Hawthorn, VIC 3122, Australia}
\author{L. J. Milligan}
\affiliation{ARC Centre of Excellence for Dark Matter Particle Physics, Australia} 
\affiliation{School of Physics, The University of Melbourne, Melbourne, VIC 3010, Australia}
\author{J. Mould}
\affiliation{ARC Centre of Excellence for Dark Matter Particle Physics, Australia} 
\affiliation{Centre for Astrophysics and Supercomputing, Swinburne University of Technology, Hawthorn, VIC 3122, Australia}
\author{F. Nuti}
\affiliation{ARC Centre of Excellence for Dark Matter Particle Physics, Australia} 
\affiliation{School of Physics, The University of Melbourne, Melbourne, VIC 3010, Australia} 
\author{F. Scutti}
\affiliation{ARC Centre of Excellence for Dark Matter Particle Physics, Australia} 
\affiliation{Centre for Astrophysics and Supercomputing, Swinburne University of Technology, Hawthorn, VIC 3122, Australia}
\author{Z. Slavkovsk\'{a}}
\author{N. J. Spinks}
\affiliation{ARC Centre of Excellence for Dark Matter Particle Physics, Australia} 
\affiliation{Department of Nuclear Physics and Accelerator Applications, The Australian National University, Canberra, ACT 2601, Australia}
\author{O. Stanley}
\affiliation{ARC Centre of Excellence for Dark Matter Particle Physics, Australia} 
\affiliation{School of Physics, The University of Melbourne, Melbourne, VIC 3010, Australia} 
\author{A. E. Stuchbery}
\affiliation{ARC Centre of Excellence for Dark Matter Particle Physics, Australia} 
\affiliation{Department of Nuclear Physics and Accelerator Applications, The Australian National University, Canberra, ACT 2601, Australia}
\author{G. N. Taylor}
\author{P. Urquijo}
\affiliation{ARC Centre of Excellence for Dark Matter Particle Physics, Australia} 
\affiliation{School of Physics, The University of Melbourne, Melbourne, VIC 3010, Australia} 
\author{A. G. Williams} 
\affiliation{Department of Physics, The University of Adelaide, Adelaide, SA 5005, Australia}
\affiliation{ARC Centre of Excellence for Dark Matter Particle Physics, Australia} 
\author{Y. Y. Zhong}
\affiliation{ARC Centre of Excellence for Dark Matter Particle Physics, Australia} 
\affiliation{Department of Nuclear Physics and Accelerator Applications, The Australian National University, Canberra, ACT 2601, Australia}
\author{M. J. Zurowski}
\affiliation{ARC Centre of Excellence for Dark Matter Particle Physics, Australia} 
\affiliation{School of Physics, The University of Melbourne, Melbourne, VIC 3010, Australia} 

\date{\today} 

\begin{abstract}

The SABRE (Sodium iodide with Active Background REjection) experiment aims to detect an annual rate modulation from dark matter interactions in ultra-high purity NaI(Tl) crystals in order to provide a model independent test of the signal observed by DAMA/LIBRA. It is made up of two separate detectors; SABRE South located at the Stawell Underground Physics Laboratory (SUPL), in regional Victoria, Australia, and SABRE North at the Laboratori Nazionali del Gran Sasso (LNGS).

 SABRE South is designed to disentangle seasonal or site-related effects from the dark matter-like modulated signal by using an active veto and muon detection system.  Ultra-high purity NaI(Tl) crystals are immersed in a linear alkyl benzene (LAB) based liquid scintillator veto, further surrounded by passive steel and polyethylene shielding and a plastic scintillator muon veto. Significant work has been undertaken to understand and mitigate the background processes, that take into account radiation from the detector materials, from both intrinsic and cosmogenic activated processes, and to understand the performance of both the crystal and veto systems.

SUPL is a newly built facility located 1024 m underground ($\sim$2900 m water equivalent) within the Stawell Gold Mine and its construction has been completed in mid-2022. The laboratory will house rare event physics searches, including the upcoming SABRE dark matter experiment, as well as measurement facilities to support low background physics experiments and applications such as radiobiology and quantum computing. 

The SABRE South commissioning is expected to occur in 2023. This paper describes the setup and projections for the experiment, and the description of the underground laboratory.

\end{abstract}


\maketitle  
\thispagestyle{fancy}  

\section{INTRODUCTION}

Dark Matter (DM) represents one of the most important open problems in modern physics \cite{bertone}. One search technique is direct detection which attempts to observe the recoil of a target after scattering with a DM particle, hypothesised as a WIMP (Weakly Interacting Massive Particles) with masses on the Gev/c$^2$-TeV/c$^2$ and cross
sections on the weak scale. This interaction rate should modulate annually, with a sinusoidal trend, due to the rotation of the Earth around the Sun as it moves through the galactic dark
matter halo. To date, only one direct detection experiment, DAMA/LIBRA, has observed a modulation compatible with that expected from DM \cite{dama}. However, this signal is in tension with every other direct detection experiment; all of which report null results for a standard WIMP \cite{tension}. These are possible explanations: the majority of experiments do not use the same target material as DAMA/LIBRA and the current experiments with sodium iodide crystals do not achieve the same sensitivity \cite{cosine,anais}. 
Further, it can be argued that the modulation observed by DAMA/LIBRA is due to seasonal effects, for example muons which also present a modulation with a maximum and minimum in the same period, in the Northern Hemisphere, as DM \cite{borex}. The SABRE Collaboration was formed to provide a model independent test of this DAMA/LIBRA signal. It foresees two detectors located in two different hemispheres: SABRE North at Gran Sasso National Laboratory (LNGS), in Italy, and SABRE South at the Stawell Underground Physics Laboratory (SUPL), in Australia. 
The two detectors have a number of common features as they are centred around the same detector module concept, and use common simulation, DAQ, and software frameworks. The two detectors differ in their shielding designs - where SABRE South will utilise a liquid scintillator system for in-situ evaluation and validation of the background to provide background rejection and particle identification. SABRE North has opted for a fully passive shielding since the organic scintillator has been phased out by LNGS. 
 The ultra-low background components will achieve the same sensitivity as DAMA/LIBRA; moreover, the site in the Southern Hemisphere will allow for the exclusion of systematic seasonal effects.

\section{The SABRE South setup}

\begin{figure}[h]
\includegraphics[width=8.8cm]{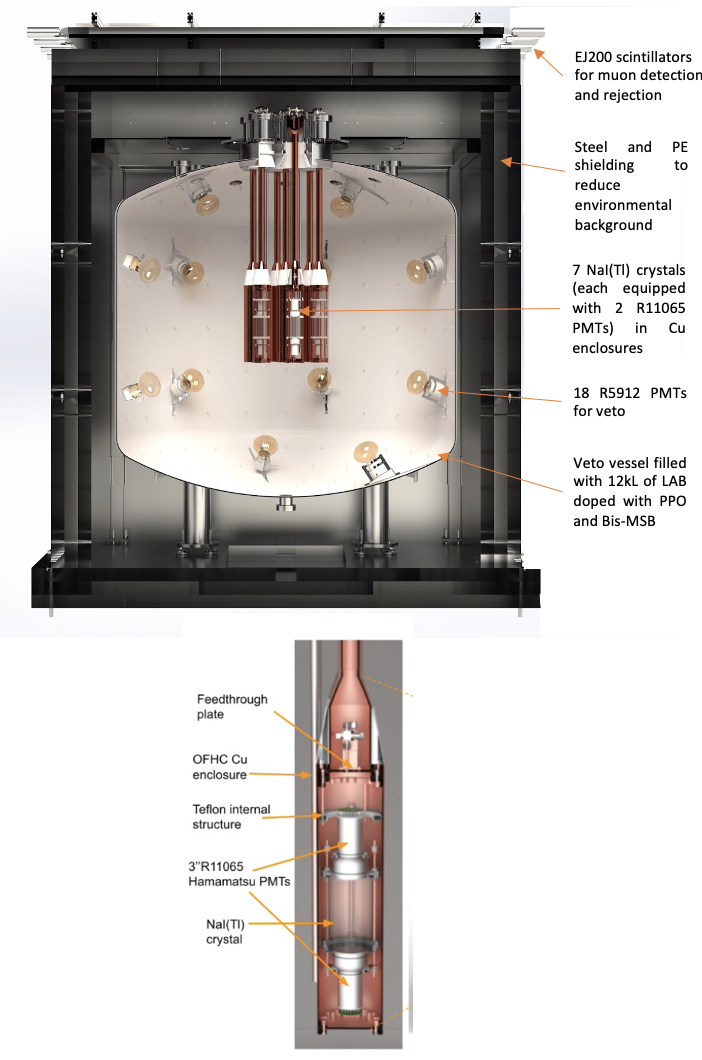}
\caption{Rendering of the SABRE South detector (top). Detail of the crystal enclosure (bottom).}
\label{layout}
\end{figure}

SABRE South is made up of three different subdetector systems: the NaI(Tl) crystal detector system, the liquid scintillator veto system, and the muon paddle detectors. The crystals and the liquid scintillator veto system are further shielded by steel and polyethelyne walls. The full experimental setup is shown in Figure \ref{layout}. The experiment can host seven NaI(Tl) cylindrical crystals 25 cm long and 5 cm in radius resulting in a mass of 7.2 kg per crystal (50.4 kg in total). Each crystal is wrapped in PTFE foil, mounted on by PTFE crystal holders, and coupled to a Hamamatsu R11065 PMT (7.6 cm diameter). Crystals are then encapsulated in cylindical oxygen-free high-thermal-conductivity copper enclosures flushed with nitrogen.

Enclosures are submersed into a liquid scintillator which is primarily used to tag and remove high energy decay products such as those from $^{40}$K. The system has a 4$\pi$ coverage, made up of 12 kL of linear alkyl benzene (sourced from JUNO) doped with PPO and Bis-MSB. Around the sides of the steel vessel this is encased in eighteen 20.4 cm Hamamatsu R5912 PMTs that are sampled at a rate of 500 MS/s. Based on preliminary optical simulations in Geant4, this system is expected to have a light yield of approximately 0.12 photoelectrons/keV, though this has a strong position dependence. With a threshold of 50 keV the inclusion of this system is able to reduce the total background by 27\%, providing a background of less than 1 cpd/kg/keV \cite{sabre_background}.\\
The muon detection system is made up of eight 3 m long EJ200 detector paddles and has a total coverage of 9.6 m$^{2}$ above the main vessel. Each paddle is coupled to a Hamamatsu R13089 PMT on each end and sampled at a rate of 3.2 GS/s. Calibrations are still ongoing, but the energy threshold of these detectors is expected to be around 1 MeV. The detectors have approximately 400 ps timing resolution resulting in 5 cm position resolution, allowing for a long term measurement of the muon flux, and particle identification when used with the liquid veto system.

\section{High-purity crystals}

Crystal radiopurity is the most important aspect of the experiment. 
Indeed, as already demonstrated by other existing NaI(Tl)-based experiments,\cite{dama,cosine,anais}, a large fraction of the background for DM search comes from residual radioactive contaminants in the crystal themselves, especially $^{40}$K and $^{210}$Pb; the latter is a daughter of the uranium decay.
Crystals are grown from Merck’s Astrograde powder, the highest purity NaI powder commercially available, which has a potassium contamination below 10 ppb, and uranium and thorium contamination
below 1 ppt. 

\begin{figure}[ht]
\includegraphics[width=8.8cm]{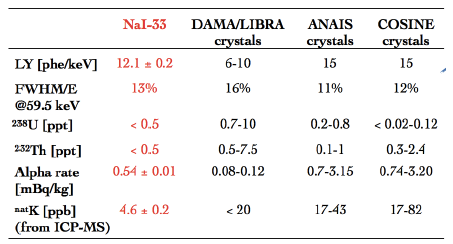}
\caption{NaI-33 crystal measurements \cite{antonello} in red compared the ones from DAMA/LIBRA \cite{damacrystal}, ANAIS \cite{anaiscrystal} and COSINE \cite{cosinecrystal}. Potassium, Thorium and Uranium concentrations are very low respect to the other NaI(Tl) experiments.}
\label{crystalTab}
\end{figure}

The SABRE collaboration, through an R\&D effort based out of Princeton in collaboration with Radiation Monitoring Devices company, has developed some of the lowest background crystals in the world with only 4.6 ppb of potassium (Figure \ref{crystalTab}). The analysed crystal mass is equal to 4.3 kg. 

RMD recently grew a NaI(Tl) crystal (NaI-35) for the SABRE South collaboration of a final mass of 3.3 kg which is currently being characterised at LNGS. Its light yield is $\sim$ 11.6 Photoelectrons/keV.

\section{Background model and sensitivity}

A full detector simulation was conducted to provide a background model for the radioactive contamination \cite{sabre_background}. 
The total radioactive background of less than 0.72 cpd/kg/keV (Table \ref{bulk_dmm_total}), dominated by NaI background (radiogenic and cosmogenic) with \textsuperscript{40}K effectively suppressed with $\sim$80\% efficiency \cite{sabre_background}. This meets SABRE South's goal of having less than 10\% of background from non-crystal sources. 
External backgrounds, shielding, LS bulk, and veto PMTs contribute less than 10\textsuperscript{-3} cpd/kg/keV.

\begin{table}[htb!]
\footnotesize
\centering
\begin{tabular}{lcc}
\hline
 & Rate & Veto Efficiency\\
 & [cpd/kg/kev]   & [\%]\\
\hline
Crystal radiogenic & $5.2 \cdot 10^{-1}$ & 13 \\
Crystal cosmogenic & $1.6 \cdot 10^{-1}$ & 40 \\
Crystal PMTs & $3.8 \cdot 10^{-2}$ & 60 \\
PTFE wrap & $4.5 \cdot 10^{-3}$ & 13 \\
Enclosures & $3.2 \cdot 10^{-3}$ & 85 \\
Conduits & $1.9 \cdot 10^{-5}$ & 96 \\
Liquid scintillator & $4.9 \cdot 10^{-8}$ & $>99$\\
Steel vessel & $1.4 \cdot 10^{-5}$ & $>99$ \\
Veto PMTs & $1.9 \cdot 10^{-5}$ & $>99$ \\
Shielding & $3.9 \cdot 10^{-6}$ & $>99$ \\
External & O($10^{-4}$) & $>99$ \\
\hline
Total & $7.2 \cdot 10^{-1}$ & 27 \\
\hline
\end{tabular}
  \caption{Background rate in the dark matter region for the SABRE South components, and the corresponding veto efficiency.}
  \label{bulk_dmm_total}
\end{table}

\begin{figure}[h]
\includegraphics[width=6.2cm]{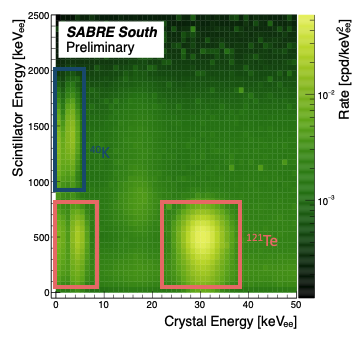}
\includegraphics[width=8.5cm]{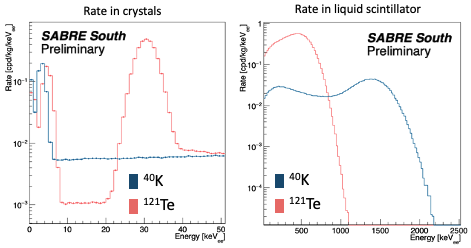}
\caption{Top: energy depositions in both the crystal and liquid scintillator. Indicated in blue and pink are events due to $^{40}$K and $^{121}$Te. Bottom left: rate recorded in crystals from $^{40}$K and $^{121}$Te. Bottom right: rate recorded in liquid scintillator from $^{40}$K and $^{121}$Te.}
\label{id}
\end{figure}

These simulations also indicated that as well as providing shielding and vetoing, the liquid scintillator system also allows for in-situ measurements of background contamination to help inform fits due to well defined correlations in the liquid scintillator and crystals, an example of which is shown in Figure \ref{id}. 

\begin{figure}[htb!]
\includegraphics[width=6.0cm]{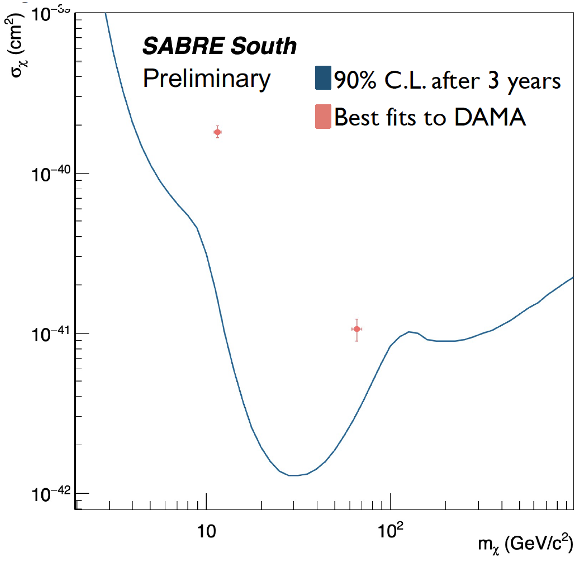}
\caption{90\% C.L. assuming a standard spin independent WIMP, a SABRE South total crystal mass of 50 kg and background of 0.72 cpd/kg/keV.}
\label{sens1}
\end{figure}

\begin{figure}[htb!]
\includegraphics[width=5.9cm]{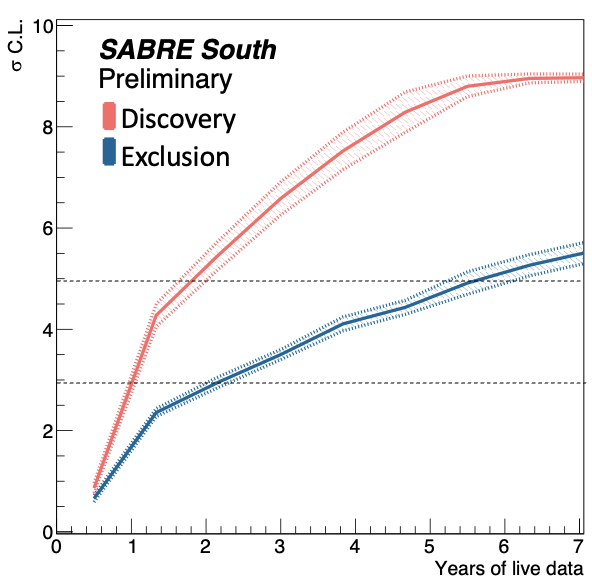}
\caption{Exclusion and discovery power of SABRE South for the DAMA/LIBRA modulation.}
\label{sens2}
\end{figure}

The projected sensitivity of SABRE South assuming this background model from simulated radioactivity and a total crystal mass of 50 kg is shown in Figures \ref{sens1} and \ref{sens2} assuming a standard spin independent WIMP and showing the evolution of discovery and exclusion power for the DAMA/LIBRA modulation. Based on these results SABRE South will be able to
refute the interpretation of the DAMA/LIBRA modulation
as a dark matter signal with 3$\sigma$ CL. In the event of observation of the annual modulation, this signal would reach a significance of 5$\sigma$ CL with two full years of data.

\section{The Stawell Underground Physics Laboratory}
\begin{figure}[h]
\includegraphics[width=8.5cm]{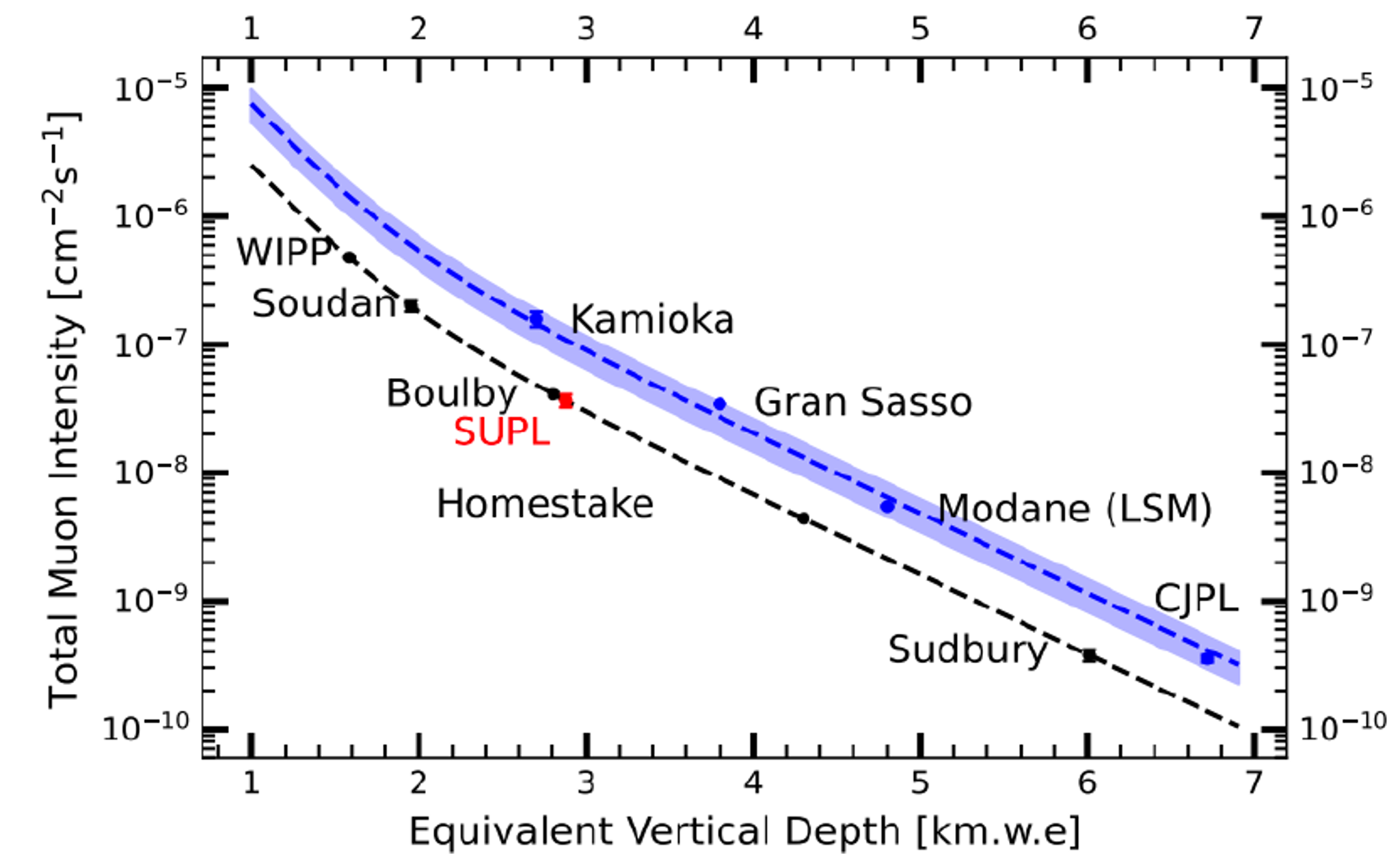}
\caption{Muon flux at a number of underground laboratories.}
\label{supl}
\end{figure}

SABRE South is to be placed at the recently completed Stawell Underground Physics Laboratory (SUPL). This is located in Western Victoria, Australia, 240 km from Melbourne. The laboratory is 1025 m below ground with a flat over burden of basalt, providing almost 3 km of water equivalent shielding, and a muon flux similar to that of Boulby, Figure \ref{supl}. \\Internal walls are pinned with steel, sprayed with low radioactivity shotcrete and coated with Tekflex to keep the radioactivity level low.\\
Construction was completed in mid-2022;  background muon, gamma, and neutron measurements are planned to start in late 2022.

\section{CONCLUSIONS}
SABRE South is part of the SABRE Collaboration designed to test DAMA/LIBRA modulation.
SABRE South is the first dark matter direct-detection experiment in the Southern Hemisphere and will be located inside the new SUPL underground laboratory.\\
High purity crystals and large active veto give ultra-low background of 0.72 cpd/kg/keV.\\
SABRE South will have 5$\sigma$ discovery (3$\sigma$ exclusion) power to a DAMA-like signal with 2 years of data taking.\\
Construction will begin later this year with commissioning expected mid-2023, after which data taking will commence.

\begin{acknowledgments}
The SABRE South program is supported by 
the Australian Government through the Australian Research Council (Grants:
CE200100008, 
LE190100196, 
LE170100162, 
LE160100080, 
DP190103123, 
DP170101675, 
LP150100705).
This research was partially supported by Australian Government Research Training Program Scholarships and Melbourne Research Scholarships.
This research was supported by The University of Melbourne’s Research Computing Services and the Petascale Campus Initiative.
We thank the SABRE North collaboration for their contribution to the SABRE South experiment design and to the simulation framework.
We also thank the Australian Nuclear Science and Technology Organisation for the assistance with the material screening and the measurement of background radiation at SUPL.
\end{acknowledgments}

\end{document}